# Software Citation Implementation Challenges

Authors: Daniel S. Katz, Daina Bouquin, Neil P. Chue Hong, Jessica Hausman, Catherine Jones, Daniel Chivvis, Tim Clark, Mercè Crosas, Stephan Druskat, Martin Fenner, Tom Gillespie, Alejandra Gonzalez-Beltran, Morane Gruenpeter, Ted Habermann, Robert Haines, Melissa Harrison, Edwin Henneken, Lorraine Hwang, Matthew B. Jones, Alastair A. Kelly, David N. Kennedy, Katrin Leinweber, Fernando Rios, Carly B. Robinson, Ilian Todorov, Mingfang Wu, Qian Zhang

Version history:
27 August 2018 - started
20 May 2019 - version 1 complete

**Table of Contents:**

# 1. Introduction

The main output of the [FORCE11 Software Citation working group](#) was a [paper on software citation principles](#) published in September 2016. This paper laid out a set of six high-level principles for software citation (importance, credit and attribution, unique identification, persistence, accessibility, and specificity) and discussed how they could be used to implement software citation in the scholarly community. In a series of talks and other activities, we have promoted software citation using these increasingly accepted principles. At the time the initial paper was published, we also provided the following (old) guidance and examples on how to make software citable, though we now realize there are unresolved problems with that guidance. The purpose of this document is to provide an explanation of current issues impacting scholarly attribution of research software, organize updated implementation guidance, and identify where best practices and solutions are still needed.

## 1.1 Document Purpose and Intended Audience

This document is not intended to support typical researchers, editors, etc., for whom we will develop and point to other slimmer, more specific primers. It is not "user documentation", for those who want to use software citation (e.g., cite the software they use, make their own software citable by others) in their writing, but "developer documentation", for those who are developing software citation practices and tools.

This document is aimed at various types of expert stakeholders (as follows) who have a level of expertise or advocacy that is above that of the average person who wishes to understand how to enact a specific software citation use case.

- **Communities that represent developers and other researchers, including software developers seeking to champion software citation** will use this document to provide more detailed and specific guidance to members of their community around using software citation
- **Publishers** will use this document to create policies and tools for software citation, including developing guidance and examples for authors, editors, and reviewers
- **Editors of journals or peer-reviewers** will use this document to develop workflows to ensure software used in publications is being cited properly
- **Librarians, archivists, and repository managers** will use this document to develop training and other resources for article authors and software developers to improve software citation practices; and advocate for or make changes to internal archival and repository systems to enable software citation.
- **Indexing and abstracting services** will use this document to develop tools and policies to expose and index software citations and associated metadata
- **Funders** will use this document to develop policies that encourage practices that enable proper citation of software developed using grant funds, as well as proper citation of software in grant proposals, and to identify unresolved gaps in tooling or methods related to software citation that may be targeted for funding.

These expert stakeholders might be those at communities, institutions or journals who support researchers, or software citation champions.

It presents a view of the current state of software citation practice and organizes updated implementation guidance to enable these stakeholders to get up to speed with the current practice of software citation, the language and terminology that is being used, and the challenges that are still to be addressed.

## 1.2 Previous guidance

When the software citation principles were published, guidance on their implementation was generally given as follows. We quickly realized that this guidance was incomplete and insufficient, however, we will use the old guidance as a starting point for understanding its limits and what new work is needed.

Old guidance to making software citable:
- Publish the software:

- If the software is on GitHub, follow steps in
  https://guides.github.com/activities/citable-code/
- If the software is not on GitHub, submit it to Zenodo directly, or to figshare, or a suitable domain repository, with appropriate metadata (including authors, title, citations, and dependencies)
- Get a DOI
- Create a CITATION file and update your README to tell people how to cite
- If needed, also write a paper about the software that discusses performance, algorithms, and other features.

Old guidance to citing software:
- Check for a CITATION file or README that contains citation information; if such a file says how to cite the software itself, do that
- If there is no CITATION file or specifications in the README, do your best to follow the principles
  - Regarding authors of the software, if the software developers declare who the authors are, list them; otherwise, just name the project (Project X for open source software, Company Y for commercial software) as the authors
  - Try to include a method for identification that is machine actionable, globally unique, interoperable – perhaps a URL to a release, a company product number
  - If there's a landing page that includes metadata, point to that, not directly to the software (i.e., if software is on GitHub and in Zenodo, point to the Zenodo version and specifically to the landing page)
  - Include specific version/release information
  - If there is a software paper that includes information important to your use of the software, possible cite that *in addition to* the citation of the software itself

## 1.3 Moving forward

Since 2016, there have been many changes in the world relevant to software citation, including the introduction of software metadata specifications such as CodeMeta and the Citation File Format, and the universal software archive Software Heritage. Groups that want to implement software citation have also asked for specific guidance on what they should do with different types of software.

Today, a series of technical and community challenges exist that make moving beyond general acceptance to implementation and common use difficult. This document attempts to list them and some thoughts on how they might be addressed.

## 1.4 Future guidance and the limits of this specific document

Once this document is completed, it will still not answer all the questions about software citation, particularly as the scholarly world and the scholarly communications community continue to change. In addition, as discussed in Section 1.1, the guidance here is not aimed at the average person who needs to enact a software citation use case. Therefore:

- Communities should build further documents based on this document, and perhaps a more end-user-oriented version of this document, which would be more specific for their communities and use cases.
  - At FORCE2018, a group that attended the Software Citation workshop started such a document.
  - The FORCE11 Software Citation Implementation Working Group has been developing specific guidance on software citation for publication authors, reviewers, and software developers.
- Later versions of this document itself will need to be written, perhaps with small changes initially, but eventually with larger changes.
- To provide feedback on this document, a version where pull requests may be made is https://github.com/force11/force11-sciwg/tree/master/Challenges/readme.md and discussion can occur through issues created in https://github.com/force11/force11-sciwg/issues

# 2. Software types

In order to cite software, it must be distinctly identified, and citation-relevant metadata, such as author names and release dates, must be gathered. There are many approaches to categorizing software, each with different metadata facets. For instance, software may be categorized by function, by layer (e.g., infrastructure, library, tool, framework, component), or by the software distribution mechanism (e.g., source code, binary executable, package, container, service). In this document, we primarily consider just three facets that are important for and where we have some clarity about software citation: the permission status, the publication status, and instantiation status. In this section we first define categories of these characteristics, and then define six types of software formed by the intersection of those categories. For each type of software, we discuss how the method for identification and the location (or absence) of metadata may differ. In Section 3 (Technological Challenges), we will cover in detail how these differences currently affect citability and citation options for these types of software according to the Principles. (Note that there remain open issues with regards to software composed of multiple layers and collections, as discussed in Section 6.12, and regarding software distribution mechanism.)

**Permission Status (e.g., License)**: Open Source, Closed Source
Software license terms are diverse, but only two general categories are needed to describe the differences relevant to citation. Software can be open source, which here includes any legal status where the source code may be viewed and reused, or closed source (aka proprietary), which includes any status that does not allow the source code to viewed and reused. (Note that explicit licenses are necessary for software to be unambiguously open, but openness might be implied by a software author's intent to allow reuse of their software; for the purpose of this document we refer to both as open.)

**Publication Status**: Published, Unpublished
"Publication" is used here to refer to a formal process of archiving a copy of the software via an entity referred to here as a publisher, and creating a resolvable identifier for it. Merely putting the software on the web is not publishing. Software is "archived" once it has been deposited with appropriate metadata in a repository (a publisher) where the curators of that repository steward the software with long-term preservation as their goal (e.g., Zenodo, figshare, institutional archival repositories). Software that has not been archived and assigned a resolvable identifier, even if it has been made public, is considered "unpublished" for the purposes of this document. Unpublished software is often publicly available, but hosted by an organization that does not commit to long term preservation (e.g., GitHub)[1].

Note that software is defined by each of these categories. Thus, in the terminology used here, we consider source code that is publicly available online to be *open source* but *unpublished*, unless it and its associated metadata have been deposited and made available via an organization that provides long-term archiving of the software and metadata, and provides identifiers for software deposits.

**Instantiation Status**: Version, Concept
Instantiation status distinguishes between concrete instances of software, called "Versions," and the abstract idea of the software project, called a "Concept", which is instantiated by its Versions. For example, the Concept software "R" is instantiated by both its first Version software "R 1.0" and its more recent Version software "R 3.5.2" (among many other versions). While in theory, a Concept can exist as an idea that has no instantiated form, for practical purposes, Concept software only exists as a citable object if it has at least one Version. Concept software can thus be generally considered a collection of Versions. (See Section 6.4 for discussion about this choice of wording.)

The intersections of these categories define six types of software as illustrated in Figure 1

---

[1] This distinction is similar to that of posting a PDF document on a web site (making it available) vs submitting it to a publisher or repository that will archive it and create a landing page and an identifier. In both cases, the document has been made available, but only in the latter case has it been published. The key difference is the intent of the document host: only short term availability or also long term archiving.

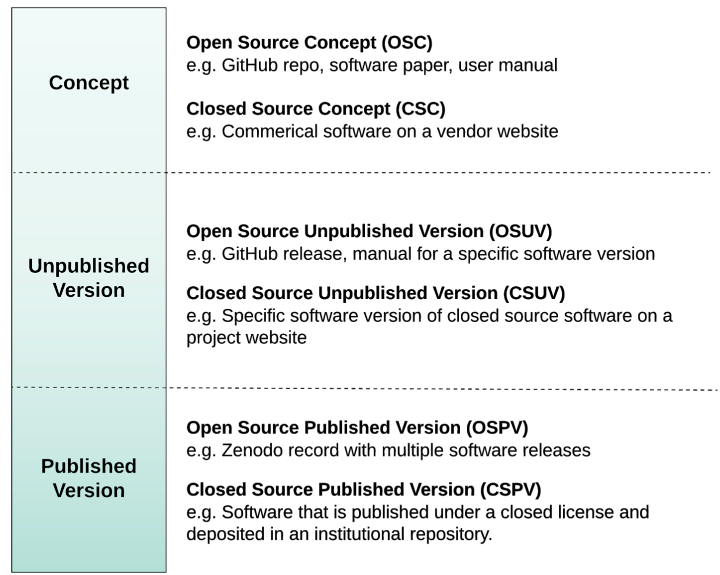

| Concept | **Open Source Concept (OSC)**<br>e.g. GitHub repo, software paper, user manual<br><br>**Closed Source Concept (CSC)**<br>e.g. Commerical software on a vendor website |
| Unpublished Version | **Open Source Unpublished Version (OSUV)**<br>e.g. GitHub release, manual for a specific software version<br><br>**Closed Source Unpublished Version (CSUV)**<br>e.g. Specific software version of closed source software on a project website |
| Published Version | **Open Source Published Version (OSPV)**<br>e.g. Zenodo record with multiple software releases<br><br>**Closed Source Published Version (CSPV)**<br>e.g. Software that is published under a closed license and deposited in an institutional repository. |

Figure 1: Differentiating between software types based on version availability

1. **Open Source Concept (OSC)** - The software concept for an open source software comprehensive package, for example, the Application Skeleton package (https://github.com/applicationskeleton/Skeleton).
2. **Closed Source Concept (CSC)** - The software concept for a closed source, possibly commercial software package, such as SAS/STAT.
3. **Open Source Unpublished Version (OSUV)** - An unpublished version of an open source software package, for example, https://github.com/applicationskeleton/Skeleton as of commit 81c66c0db5c381dacc0841a4c16e0b3876b15b89.
4. **Closed Source Unpublished Version (CSUV)** - An unpublished version of closed source, possibly commercial, software. An example is version 14.3 of SAS/STAT.
5. **Open Source Published Version (OSPV)** - A published version of an open source software package. An example is version 1.2 of the Application Skeleton package (https://github.com/applicationskeleton/Skeleton), found on GitHub at https://github.com/applicationskeleton/Skeleton/releases/tag/v1.2, published by Zenodo as https://doi.org/10.5281/zenodo.13750.
6. **Closed Source Published Version (CSPV)** - A published version of a closed source software package. Metadata may be publicly accessible, but software may be embargoed or only accessible internally to the institution. For example, software that is published under a closed license and deposited in an institutional repository.

Note that strictly speaking, a Concept cannot have a license, as it's an abstract idea; only an instantiation of a Concept (a Version) can have a license. Similarly, Concepts cannot strictly be said to be Published, since the definition of Publication requires code to be archived. Once again, written code that could be archived has been instantiated, so Publication status can only be given to Versions. However, for practical purposes, Concepts have the Permission and Publication statuses of their Versions, and for simplicity in these six types, Permission (e.g. license) and Publication statuses are assumed to be consistent between Versions. However, it is certainly possible that a software project might switch between open-source Versions and closed-source Versions, for example, and thus have a Concept that encompasses both, which would not fit neatly into this typology. The discussion on Hybrid Types below highlights some other such instances and the implications for citation.

**Hybrid Types**

In practice, a piece of software does not always fit neatly into a single type; software may be composed of both open source and closed source components, or even of both published and unpublished components. In this case, it is important to be able identify each file or component uniquely and independently so that it is clear what metadata and software type apply to each component: license, version, creators, etc. There may also be hybrids of software and other types of objects, such as a Jupyter notebook or a computational workflow or script that is published within mixed-type objects known as Research Objects or Research Compendia that represent computational pipelines with their associated provenance. An example of a case is the Chlorophyll R Script (software) contained in the mixed-type Gentry et al. research data package (https://doi.org/10.5063/F1CF9N69); this package, including but not limited to the software, is cited in an associated paper (https://doi.org/10.1038/s41559-017-0257-9). Another example exists in the data archive at Johns Hopkins University, where a record contains both the data and software associated with a research article. Similarly, computational pipelines and frameworks may be collections of elements of the above types. In these cases, the specific components used may vary when the software is used and who maintains different components of the pipeline may change overtime. For example spec2d (https://www2.keck.hawaii.edu/inst/deimos/pipeline.html), a spectral data reduction pipeline, is based on the proprietary language IDL and was developed by a team at the University of California, Berkeley using NSF funding. The spec2d pipeline is based on the Sloan Digital Sky Survey spectral reduction package written by a different group, and the core IDL utilities (a separate package) are maintained by a team at Princeton. And both IDL packages are needed to use the DEIMOS data reduction pipeline. How to cite these objects is still unclear - see Section 6.12.

# 3. Fundamental challenges

The primary technical challenges to software citation implementation are how to identify software, how to cite it once it has been identified, and how citations are counted.

At a high level, the software citation principles and discussion essentially tell us how to cite open source software versions: they say that versions of open source software should be published, and that those versions (and their corresponding DOIs) should then be cited. Today, with Software Heritage, we also have the possibility of citing unpublished versions (though Software

Heritage doesn't store any of the metadata that we would like, unless it has been stored in the repository by the software authors) of open source software, by pointing to an archived copy of a specific commit or release in the code repository. We do not currently have a generally accepted solution for versions of closed source software, or software concepts, whether open or closed source.

## 3.1 How to identify software

How to identify software to be cited is an open question that has been discussed in the FORCE11 software citation working groups, and will be discussed in the [RDA Software Source Code Identification Working Group](#).

1. OSC - The software can be identified by the software repository, and is sometimes represented by a software paper or the software users manual. A persistent identifier may be created or found in a registry such as [SciCrunch](#) or [ASCL](#).

   While the idea of an identifier (such as a DOI) has always been possible to associate with software concept, there is currently no public way to directly obtain such an identifier (see Section 6.8). If there was such a way to do so, as of DataCite schema 4.1, the relation types `hasVersion` and `isVersionOf` would allow us to link `concept DOIs` and `version DOIs`. Some indirect options include

   a. A version of the software is published to Zenodo, a concept identifier (see Section 6.13) is created. Multiple versions of the software, each with their own identifier, are then published, and link to the concept, using the relation types.

   b. While some in life sciences might use an RRID to identify a software concept, this type of identifier is unknown outside life sciences, and is not intended to identify concepts alone, but rather to indicate the use of an executable form of software. The underlying SRC identifiers could be thought of as identifying a software concept, and the RRID metadata for digital resources has been [aligned](#) with the DataCite relation types.

   c. Organizations who register DataCite DOIs can take advantage of the built-in support for codemeta. An example of this is [https://doi.org/10.5438/qeg0-3gm3](https://doi.org/10.5438/qeg0-3gm3), which points to [https://github.com/datacite/maremma](https://github.com/datacite/maremma). This DOI used the GitHub URL as metadata. When the DOI was created, the DataCite DOI registration service fetched the codemeta.json from the repository, converted the metadata into DataCite XML, and stored them.

2. CSC - Perhaps this is a URL to the software, or a company product number. This could simply be a product name, such as SAS/STAT. A persistent identifier may be created or found in a registry such as [SciCrunch](#) or [ASCL](#).

3. OSUV - If the software has been archived by [Software Heritage](#) (they have archived a large amount of open source software and are working towards doing this for all open source software). Their [search and browse capability](#) can be used to find a version of the software to be cited. If the repository is not yet archived by Software Heritage, the [Save](#) [Code Now](#) feature can be used to request that a repository be archived, using its URL[2]. The permalinks button on the right to retrieve the persistent identifier (or the URL). For example, [https://github.com/applicationskeleton/Skeleton](https://github.com/applicationskeleton/Skeleton) as of commit 81c66c0db5c381dacc0841a4c16e0b3876b15b89 is [https://archive.softwareheritage.org/swh:1:rev:81c66c0db5c381dacc0841a4c16e0b3876b15b89;origin=https://github.com/applicationskeleton/Skeleton/](https://archive.softwareheritage.org/swh:1:rev:81c66c0db5c381dacc0841a4c16e0b3876b15b89;origin=https://github.com/applicationskeleton/Skeleton/) in Software Heritage's archive. Metaresolvers can also be used here, for example, this is [http://n2t.net/swh:1:rev:81c66c0db5c381dacc0841a4c16e0b3876b15b89;origin=https://github.com/applicationskeleton/Skeleton/](http://n2t.net/swh:1:rev:81c66c0db5c381dacc0841a4c16e0b3876b15b89;origin=https://github.com/applicationskeleton/Skeleton/) and [https://identifiers.org/swh:1:rev:81c66c0db5c381dacc0841a4c16e0b3876b15b89](https://identifiers.org/swh:1:rev:81c66c0db5c381dacc0841a4c16e0b3876b15b89) using the [n2t.net](http://n2t.net) and [identifiers.org](http://identifiers.org) metaresolvers. (Note that the identifiers.org version does not currently allow the origin to be added.)

4. CSUV - Perhaps this is a URL to the software, or a company product number. This could simply be a product name, such as SAS/STAT version 14.3. A persistent identifier may be created or found in a registry such as [SciCrunch](#) or [ASCL](#).
5. OSPV - The software is identified by the publisher, likely with a DOI. For example, version 1.2 of the Application Skeleton package published by Zenodo is [https://doi.org/10.5281/zenodo.13750](https://doi.org/10.5281/zenodo.13750).
6. CSPV - Perhaps this is a URL to a release, or a company product number, or simply a product name and version number. For example, SAS/STAT ver. 14.3

## 3.2 Metadata for software citation

The generally required metadata for software citation from the [Software Citation Principles](#), are listed as follows:

● Identifier - Applicable to all; persistent identifiers preferred. Versionless DOIs can be used for CSC and OSC to identify software as a concept rather than a specific version/release, though there are not generally available tools to generate/issue such DOIs at this time
● Software name - Applicable to all
● Authors - Applicable to all; if not able to be determined, one can fall back to "[Software name] Project" for open source projects, or "[Entity name]" or "[Company name]" commercial projects
● Version/Release date - for CSC and OSC, likely the date of the first version, or possibly the range of dates for all versions

---

[2] The timing of such archiving may vary based on:
1. The URL provided. If it is from a known domain (github for example) it will be accepted directly, while an unknown URL domain will have to pass a manual review, and the time needed for this depends on human resources at the time of submission.
2. The size of the repository. The number of commits can make a big difference in the ingestion time.
3. The maintenance state of the Software Heritage infrastructure
Archiving a normal size repository can take an hour, but this is not guaranteed. You can follow the request status on the "browse save requests" tab.

- Location/Repository of working source code - May not be directly applicable to closed source, though some closed source could provide a location that is not accessible, so possible values of this metadata should be: location or not available

This is also shown in Figure 2.

## 3.3 Storing metadata

Metadata for software citation may be stored and made machine actionable in one or more of three places, or may not be stored at all. These are: 1) the metadata associated with an identifier (e.g., a DOI[3]), 2) in CodeMeta.json in a software repository alongside the code to be cited, and 3) in CITATION.cff files in a software repository alongside the code to be cited.

Note that both CodeMeta and CFF can be used to provide citation-relevant metadata for a software and may both be present. The primary difference between CFF and CodeMeta is that CFF is a "front-end" format, and CodeMeta is a "back-end" format. See Section 3.3.1 for more details and a comparison.

The following table shows the different types of software and summarizes where metadata describing them can be stored.

| Type | No stored metadata | No publically accessible metadata[4] | DOI metadata | CodeMeta metadata | CFF metadata |
|------|--------------------|--------------------------------------|--------------|-------------------|--------------|
| OSC  |   |   | X (if published) | X | X |
| CSC  | X | X |   |   |   |
| OSUV |   |   |   | X | X |
| CSUV | X | X |   |   |   |
| OSPV |   |   | X | X | X |
| CSPV |   |   | X | X | X |

In summary, metadata is generally not stored or is stored inaccessibly for CSC and CSUV, though it can be stored externally in public software registries, such as SciCrunch and ASCL. (See Section 6.11.) Metadata can be stored externally via the DOI system for OSPV, CSPV, and OSC, and stored internally as a CFF or CodeMeta file for OSPV, CSPV, OSUV, and OSC.

---

[3] Version 4.1 of the DataCite schema includes the metadata needed for software citation as defined in Smith et al.

[4] Examples of public registries that store metadata for closed source software include SciCrunch and ASCL. (See Section 6.11.)

Figure 2 builds on Figure 1, describing the different types of software (from Section 2), and adds information about metadata (from Section 3.2) and where it is stored (from Section 3.3).

| Metadata<br>• Identifier<br>• Software name<br>• Software authors<br>• License<br>• Version/Release date<br>• Location/Repository | | **Closed Source**<br>If authors cannot be determined use<br>["Entity Name"]; the "entity" may be a company | **Open Source**<br>If authors cannot be determined use<br>["Software Name] Project" |
|---|---|---|---|
| | **Concept**<br>• No resolvable identifier | **Closed Source Concept (CSC)**<br>e.g. Commerical software on a vendor website<br><br>Metadata is not stored but is publically accessible | **Open Source Concept (OSC)**<br>e.g. GitHub repo, software paper, user manual<br><br>Metadata is stored as CFF and/or CodeMeta |
| | **Unpublished Version**<br>• No resolvable identifier | **Closed Source Unpublished Version (CSUV)**<br>e.g. Specific software version of closed source software on a project website<br><br>Metadata is not stored but is publically accessible<br>Cite version | **Open Source Unpublished Version (OSUV)**<br>e.g. GitHub release, manual for a specific software version<br><br>Metadata is stored as CFF and/or CodeMeta<br>Cite version or a specific commit via Software Heritage |
| | **Published Version**<br>• Resolvable identifier | **Closed Source Published Version (CSPV)**<br>e.g. Software that is published under a closed license and deposited in an institutional repository.<br><br>Metadata is stored as DOI, CFF and/or CodeMeta<br>Cite version | **Open Source Published Version (OSPV)**<br>e.g. Zenodo record with multiple software releases<br><br>Metadata is stored as DOI, CFF and/or CodeMeta<br>Cite version |

Figure 2: Metadata recommendations specified by software type and licensing information

Unresolved Issues:
- When there is no stored metadata, what should people wanting to cite the software do to meet the software citation principles? See discussion in Section 6.11.
- When there is more than one set of metadata stored, which is primary? It is important for discovery that the metadata about the object describes the type of object. Object typing for software is a developing field and may be addressed in the RDA Software Source Code Identification Working Group.
- Note that R has guidance already, and that guidance does not match the software citation principles. For instance the guidance provided by the R Project does not include a version number or license information.
  https://cran.r-project.org/doc/manuals/R-exts.html#CITATION-files
- There is a recommended citation method for bioconductor that now assigns DOIs to code.

Also see Section 6.3.

### 3.3.1 CodeMeta and the Citation File Format (CFF) comparison

CodeMeta is a machine-actionable, general-purpose exchange format for software metadata expressed in JSON-LD format (i.e. JavaScript Object Notation for Linking Data). The use of JSON-LD in the CodeMeta specification means that the software metadata includes some structure (in JSON format) and can be extended with semantics via context files, which provide mappings of the elements to vocabularies. The provided context in CodeMeta relies mostly on the schema.org[5] vocabulary. CodeMeta provides a common vocabulary and crosswalks to software metadata concepts from 25 common software metadata dialects (https://codemeta.github.io/crosswalk/), which provides the ability to automate the conversion of software among these 25 dialects.

The Citation File Format (CFF) is a machine-actionable, human-readable and -writable metadata format based on the Software Citation Principles. CFF relies on the YAML format for representing the citation metadata, which is a superset of JSON (i.e., JavaScript Object Notation). It requires at least the minimal set of information for a software to conform to the Software Citation Principles, and allows for scoped references, e.g., for references only relevant for a part of the complete source code of a software, or references that are relevant only when the software is run with a specific configuration.

While CFF is a suitable format for the initial provision of software citation metadata by the creators of a software, the metadata it provides should also be transferred to CodeMeta downstream in the software citation workflow. Transferring the CFF metadata lets software authors leverage CodeMeta's linked-data features and can enable more comprehensive documentation of the software metadata beyond bibliographic elements.

A good example (circa 2018) for how the two formats can be used together is the Netherlands eScience Center's Research Software Directory (e.g., https://research-software.nl/software/xenon), where CFF is used to provide the actual citation metadata input (converted to BibTeX etc.), and is also used to generate CodeMeta JSON-LD using cffconvert, which is embedded in a directory entry's landing page where it can be located by search engines.

## 3.4 Converting citation metadata

Once citation metadata is known and stored, it can be converted to various machine- and human-readable formats, such as a text citation, or to a format such as bibtex, RIS, etc. This metadata can also be stored in a reference manager, which can then do this conversion.

For example, DataCite's Content Resolver can return a representation of DOI in different formats. An example of this is returning the metadata for version 1.2 of the Application Skeleton package (with doi https://doi.org/10.5281/zenodo.13750) in bibtex:

```
> curl https://data.datacite.org/application/x-bibtex/10.5281/zenodo.13750

@misc{https://doi.org/10.5281/zenodo.13750,
```

---

```
 doi = {10.5281/zenodo.13750},
 url = {https://zenodo.org/record/13750},
 author = {Katz, Daniel S. and Merzky, Andre and Turilli, Matteo and Wilde,
Michael and Zhang, Zhao},
 keywords = {computer science, application skeleton, co-design, distributed
computing, many-task computing, parallel computing},
 title = {Application Skeleton V1.2},
 publisher = {Zenodo},
 year = {2015}
}
```

Similarly, citation metadata provided in the Citation File Format can be converted to different formats such as a CodeMeta representation in JSON-LD, BibTeX, RIS, EndNote and Zenodo JSON with a tool such as [cffconvert](). An example of this is converting the current *CITATION.cff* file for [nlppln]() to BibTeX and CodeMeta JSON-LD. The below examples show the local use scenario where *cffconvert* is run locally in the folder containing the metadata. Using a *cffconvert* webservice (for BibTeX:

[https://us-central1-citation-cff.cloudfunctions.net/convert?url=https://github.com/nlppln/nlppln&format=bibtex](); for CodeMeta:

[https://us-central1-citation-cff.cloudfunctions.net/convert?url=https://github.com/nlppln/nlppln&format=codemeta]()) yields the same results.

```
> cffconvert -f bibtex

@misc{YourReferenceHere,
author = {
        Janneke M. van der Zwaan and
        Dafne van Kuppevelt
        },
title = {NLP Pipeline (nlppln)},
month = {1},
year = {2019},
doi   = {10.5281/zenodo.1116323},
url   = {https://github.com/nlppln/nlppln}
}

> cffconvert -f codemeta

{
        "@context": [
        "https://doi.org/10.5063/schema/codemeta-2.0",
        "http://schema.org"
        ],
        "@type": "SoftwareSourceCode",
        "author": [
        {
        "@id": "0000-0002-8329-7000",
```

```
    "@type": "Person",
    "affiliation": {
    "@type": "Organization",
    "legalName": "Netherlands eScience Center"
    },
    "familyName": "van der Zwaan",
    "givenName": "Janneke M."
    },
    {
    "@type": "Person",
    "affiliation": {
    "@type": "Organization",
    "legalName": "Netherlands eScience Center"
    },
    "familyName": "van Kuppevelt",
    "givenName": "Dafne"
    }
    ],
    "codeRepository": "https://github.com/nlppln/nlppln",
    "datePublished": "2019-01-08",
    "identifier": "https://doi.org/10.5281/zenodo.1116323",
    "license": "http://www.apache.org/licenses/LICENSE-2.0",
    "name": "NLP Pipeline (nlppln)",
    "version": "0.3.3"
}
```

Unresolved Issues

Note that the reference entries above are "@misc" type and not "@software". Although the @software entry type is supported by the standard BibTeX data model, only a handful of custom styles, such as [biblatex-oxref](#) and [biblatex-apa](#) have dedicated @software bibliography drivers. The lack of such drivers in most standard styles, such as Biber, leads to @software types defaulting back to the @misc driver.

Soft mapping @software to @misc is problematic for several reasons. Mainly, there is no version field in @misc, and this often leads to developers including version numbers in title fields (e.g., "Title (Version {version})"). Similarly, "software" is added as a keyword type field in @misc entries (e.g, " type = {Computer software}" or "keywords = {type:software}"). Although some styles, such as BiblaTeX, automatically translate custom software keywords accordingly, most styles will subsequently mix software citations with other non-software @misc entry types in output. The issue is that custom keyword fields are designed primarily for filtering and not for output, posing significant challenges to reference management tools like EndNote and Zotero when exporting bibliographies with numerous entry-types.

A second issue is that indexers need to be able to ingest CFF and CodeMeta files to properly record metadata for software entries.

## 3.5 How to cite software in text

How to cite identified software is also a challenge. Here we provide a set of guidelines for the author of a paper who wants to cite software used in that paper:

1. OSC - Authors need to do their best at gathering the metadata for a citation and formatting it appropriately. If there is already metadata stored in a software registry, it can be used (or updated if needed). And if not, the author should submit the gathered metadata to a registry. (See Section 6.11.)

2. CSC - Authors need to do their best at gathering the metadata for a citation and formatting it appropriately. If there is already metadata stored in a software registry, it can be used (or updated if needed). And if not, the author should submit the gathered metadata to a registry. (See Section 6.11.)

3. OSUV - For example, for https://github.com/applicationskeleton/Skeleton as of commit 81c66c0db5c381dacc0841a4c16e0b3876b15b89, I first look for the metadata in the archive of the repository. In this case, there is an AUTHORS file that tells me the authors of the software. If this wasn't here, I would name the authors as "Application Skeleton Project." To cite this version, I might use (in a more or less IEEE style):
   Daniel S. Katz and Andre Merzky and Matteo Turilli and Michael Wilde and Zhao Zhang, "Skeleton," [Software source code], commit 81c66c0db5c381dacc0841a4c16e0b3876b15b89, 2016.
   https://archive.softwareheritage.org/swh:1:rev:81c66c0db5c381dacc0841a4c16e0b3876b15b89;origin=https://github.com/applicationskeleton/Skeleton/
   (Or we could use the compact identifiers version:
   https://identifiers.org/swh:1:rev:81c66c0db5c381dacc0841a4c16e0b3876b15b89
   If the software has been previously versioned, use the previous version's citation information as a starting point, adding to it any updates made to the AUTHORS, CITATION, README, ACKNOWLEDGEMENT etc. files as appropriate, along with the short commit hash.

4. CSUV - Authors need to do their best at gathering the metadata for a citation and formatting it appropriately. If there is already metadata stored in a software registry, it can be used (or updated if needed). And if not, the author should submit the gathered metadata to a registry. (See Section 6.11.)

5. OSPV - the metadata we need for citation should be available as part of its publication, and we can cite the software similarly to how we would cite any other published digital object, with the exceptions of identifying the version (much as you would identify volume, issue, or page numbers for a journal publication.) In the specific example (https://doi.org/10.5281/zenodo.13750) we have a landing page associated with the DOI that provides a suggested citation:
   Daniel S. Katz, Andre Merzky, Matteo Turilli, Michael Wilde, & Zhao Zhang. (2015,

January 5). Application Skeleton v1.2. Zenodo. http://doi.org/10.5281/zenodo.13750
This is not perfect, since it is not marked as software, and the version is not listed, except in the title, but it's fairly easy to start with this and fix it. We could also use https://citation.crosscite.org/ to get a citation in various styles, such as IEEE, which gives:
D. S. Katz, A. Merzky, M. Turilli, M. Wilde, and Z. Zhang, "Application Skeleton V1.2." Zenodo, 05-Jan-2015.
An unresolved issue is that most citation styles currently do not indicate that a citation is software, as the Software Citation Principles recommends, and similarly, may not have a way of indicating the software's version. This is not an issue for the author of a paper citing software, but is an issue for publishers, who may need to change their citation styles. (See Section 6.7.)

6. CSPV - Authors need to do their best at gathering the metadata for a citation and formatting it appropriately. If there is already metadata stored in a software registry, it can be used (or updated if needed). And if not, the author should submit the gathered metadata to a registry. (See Section 6.11.)

All software citations should be included in the references section of papers rather than in acknowledgments or footnotes to ensure citations can be properly indexed, as stated in the software citation principles paper.

# 4. Stakeholder challenges

In order to implement software citation, in addition to the technical challenges in the previous section, we also have to bring this into the scholarly culture. We see at least four ways of looking at and interacting with the scholarly community. These can generally be viewed as points of leverage, at which we can make changes that will affect the overall community.

## 4.1 Disciplinary communities

One way of implementing software citation is by working with disciplinary communities (e.g., astronomy), which might involve professional societies (e.g., AAS), publishers (e.g., AAS again, and also see the next subsection), archives (e.g. Zenodo), indices (e.g., ADS), and other organizations.

To impact these communities, we likely need representatives for each, and those representatives should form a group to share their experiences, successes, lessons, and challenges with each other.

Actions for disciplinary communities include:
● Providing guidance for how software is cited in publications and events organized by the community, oriented towards editors, organizers, authors, and reviewers.
● Providing guidance for how software is recognized and counted within discipline

- Providing guidance for what approaches to provide software citation metadata are recommended on submission to publications and events, oriented towards authors and reviewers
- Guidance and resources to help disciplinary communities advocate for other stakeholders to adopt best practices

## 4.2 Publishers

Similar to disciplinary communities, each publisher has their own systems, practices, and culture, but also share a larger culture.

To impact these publishers, we likely need a representative for each, and those representatives should form a group to share their experiences, successes, lessons, and challenges with each other.

Actions for publishers include:
- Promote the need for authors to provide links to published copies of their code and determine how publishers can identify these
- Determine how software is internally identified in publisher systems now and how this could be improved
- Decide whether software associated with the publication (software developed by the authors of the publication) and software that is reused (software not developed by the authors of the publication, or not developed for work described in the publication) should be differentiated from each other
- Determine what policies should be adopted to support software citation
- Provide guidance and training for how software should be cited in publications to authors, reviewers, and editors; enforce these guidelines
- Ensure that publication workflows do not degrade metadata required for software citation
- Provide discipline-specific examples of software-related publisher policies, with annotations to explain why they are well-made and the important elements they include
- Provide guidance on repositories that meet criteria for software deposition

## 4.3 Repositories

Repositories archive both data and software from computational workflows, often in integrated packages that document processes used in scholarly and other work. These integrated packages often contain software, but that software is typically not specifically designed for reuse. Rather, it implements the particular computations needed for an analysis or project, and it in turn depends on all of the types of software described in Section 2. Challenges surrounding these mixed data and software packages include:
- Embedding of software (e.g. scripts, Jupyter notebooks) with data inputs, outputs, and other computation artifacts in mixed packages that are assigned a single DOI in which the software may not be individually referenceable; perhaps a separate DOIs should be created so these components can be versioned independently and linked to directly, but this is still ongoing work.
- Metadata for packages derives primarily from data repository communities but includes sections on software used, and these repositories do not yet embrace CodeMeta and similar efforts
- Need for documenting the provenance relationships between software, data, and products that describes the lineage of computational workflows
- Software in repository packages that import other software packages (e.g., via implicit language-specific library imports) but does not document these dependencies in metadata

Another challenge is related to the production or requirement for landing pages for software, and how these should be verified when software is submitted to a repository. Currently, there are only a few repositories that support software deposit or publishing. To achieve wide adoption of software citation, more repositories should support software publishing, which should include support for a landing page with human and machine-readable software metadata, a clear software citation with a globally unique and persistent identifier, and information on how to access the archived software package or code itself, and ideally also the "live" software in its development repository. Software Repositories become the "publishers" of each software version, in a similar way that repositories have become the "publishers" of datasets and guarantee the the citation will always resolve to a landing page.

## 4.4 Indexers

Indexers such as Google, Elsevier, Smithsonian Astrophysical Observatory (SAO) and NASA, etc. have created and maintain indices such as Google Scholar, Web of Science, the Astrophysics Data System (ADS).

These indices need to:
- Indicate that stored software is software (store it as a software type). Note that this may require requesting this metadata from the publisher of the software, e.g., author names.
- Capture software mentions in publications, etc.
- Provide metrics across software versions
- Develop and pilot tactics for determining metrics across different representations of software

### 4.4.1 How to count citations

This is a complex problem, with an interlinked set of challenges to be solved:
- It is necessary to identify which citations in reference lists are software
  - Perhaps by dereferencing of DOI metadata
- Counting citations to software which do not have PIDs with adequate associated metadata will complicate this issue
- How to add citations of a concept across versions is unclear
  - This could require advances in tooling

- ○ And work by indexing services
- There is a possible concern with identifying repeated citations to software in different locations within a publication (e.g. citation within a figure caption and a citation in a reference list)
  - ○ We need to ensure any particular software is only counted once.
  - ○ Given that we are promoting (and counting) formal citations for software (in the references section), this should not be an issue.
- Another issue is author disambiguation, where in some cases authors may be listed by their names, and in other cases, by their usernames (e.g. GitHub handle).

## 4.5 Funders

As the organizations that support much of the work that leads to research software, funders have a large part to play in the collection of information relating to software production, particularly through research impact assessment platforms (e.g., ResearchFish). They may also mandate policies (e.g. data management plans) that may steer communities towards best practice in software citation, for instance to make it clear how software citations should be assessed in peer review. Effective grantee requirements related to software should address how the policy applies to multiple versions, the expected time frame for compliance, consequences for non-compliance, and circumstances that permit exceptions.

Actions for funders include:
- Determine how best to reference software in proposals, research outcomes submissions and research output management plans
- Provide additional guidance about submitting software as research outputs to funders, especially focused on how software should be submitted in the case where a DOI is assigned by the funder
- Define requirements or expectations for depositing grant-funded software source code and citation-relevant metadata in a preservation archive that issues software DOIs, thus making it fully citable
- Provide guidance on repositories that meet criteria for software deposition

## 4.6 Institutions

Institutions, as the organization that generally employ researchers, have a large amount of influence on the community culture. They have a large part to play in the collection of information relating to software production in particular through current research information systems (CRIS) such as PURE, Converis or Symplectic Elements. (While institutions also may manage repositories, this topic has already been discussion in Section 4.3.)

Actions for institutions include:
- Determine what elements are necessary in software citation to measure researchers' output, including their software contributions (multi-dimensional as software is not just code), to support hiring and promotion decisions.

- Provide guidance on how to record software as a research outcome or impact in their CRIS. (For example, https://code4ref.github.io/ is an ongoing project aiming to provide such guidance for CRIS used by UK universities, currently covering Pure and RIS systems.)
- Incorporate software citation principles and metadata standards into their institutional repository workflows and metadata schemas

# 5. Previous thoughts on challenges

Note that some of us also were thinking about this in 2016, when we published "The Challenge and Promise of Software Citation for Credit, Identification, Discovery, and Reuse". In that paper, we listed the following challenges — and possible solutions/methods.

- Identify necessary metadata associated with software for citation — We suggest metadata here, and the CodeMeta project is working to determine minimal metadata.
- Standardize proper formats for citing software in publications — The FORCE11 Software Citation Working Group is defining, and gaining community acceptance for, software citation principles, after which a follow-on group will begin implementation efforts.
- Establish mechanisms for software to cite other software (i.e., dependencies) — Software publication as software papers allows this. For software that is directly published, this is an open challenge, though the fact that dependencies can be extracted from build scripts may help.
- Develop infrastructure to support indexing of software citations within the existing publication citation ecosystem — The FORCE11 Software Citation Implementation Working Group will also work on this in collaboration with publishers and indexers.
- Determine standard practices for peer review of software — Professional societies and science communities need to determine how this will happen.
- Increase cultural acceptance of the concept of software as a scholarly product) — Acceptance will happen over time; unclear how to accelerate this process.

# 6. Discussion topics (based on items and comments above)

## 6.1. Goal of this document/process

This document was produced over about a six month period at the end of 2018 and beginning of 2019, with the goal of capturing the state of software citation and identifying challenges that need to be resolved and other next steps. The FORCE11 Software Citation Implementation Working Group could try to address these challenges, or could promote them to others who want to address them. Overall, this document can be thought of as an addendum to the Software Citation Principles, discussing issues that have come up in moving from stating principles to implementing them.

## 6.2. Repositories

Researchers seeking to archive software expect repositories to provide documentation explaining how to describe software deposits so that software citation is fully supported. However, this expectation is not necessarily met by repositories that currently accept software deposits.

A challenge for repositories is how metadata is collected, for example, a repository may recommend depositors create a CITATION or CodeMeta file for their submission, or might use a form specific to software deposits. In general, we believe that software depositors should follow the Software Sustainability Institute's [guidance for software deposits](#) and ensure that all metadata elements required for citation are documented in the repository record, but this needs to be determined by each repository. Another challenges is what functionality repositories should provide to support software citation?

A challenge for software depositors is which repositories should be used for software. The FORCE11 Software Citation Implementation Working Group might provide guidelines on what the characteristics of a repository should be (to guide people in choosing say a disciplinary specific vs. general use repository), and an appendix with a list of examples.

## 6.3. Citation files

We have multiple types of citation files now: codemeta.json, citation.cff, CITATION files as proposed/used by CRAN. What guidance can we provide users, if any? Can we provide tools or methods for translation between these files?

- An example is cffconvert ([https://www.research-software.nl/software/cffconvert](https://www.research-software.nl/software/cffconvert)), which can convert CFF to BibTeX, EndNote, RIS, codemeta, plain JSON, Zenodo JSON.

## 6.4. "Concept" language

We have chosen to use the word "concept" in this document to refer to the generic idea of a piece of software. This is distinct from a version of the concept. For example, Microsoft Word is a concept, and Microsoft Word 13 and Microsoft Word 16.0.1 are both versions. We are aware that there is other language used in other communities, such as "work" (or "creative work"), "product", "software family", and "project", but given that there is no consensus for the appropriate work in the context of software, we have chosen to use "concept".

Here, we use this to distinguish between the abstract concept vs the specific representations (in this case in the form of software versions). This is similar to the dichotomy in datasets, where there is the conceptual dataset and the specific representation (in W3C DCAT - and schema.org as is based in DCAT, as well as in DATS) this is Dataset vs Distribution). In REST, there is the distinction between resource and representation.

Note: In library science, the difference between a conceptual project and its specific instantiations has been addressed in a concept model called FRBR (Functional Requirements for Bibliographic Records). While the specific language of FRBR might be confusing to non-librarians, it is mentioned here for librarians. The model includes work, expression, manifestation, and item. The first two probably address the issue here: A "work" is the conceptual level of a creation. For example, the novel "Gone With the Wind". This would correlate to "concept" as used here. Microsoft Word 2016 would be a software work. An "expression" is what we might call an edition of that work. For example, the original text and a translated text of "Gone With the Wind" would both be considered expressions. Updated editions that do things like correct typos and modernize spelling would also be considered expressions. This would probably correspond to versions or releases of software. Microsoft Word 2016 for Windows 32 bit version 16.16 would be an expression.

"[A Framework for Software Preservation](#)" follows the FRBR model, but uses the terms product, version, variant, and instance in place of work, expression, manifestation, and item.

## 6.5. Non-developer identification of software

Based on discussion in Section 3.1 and 3.2 and 3.3 and 3.5.

Here we are distinguishing between situations where software creators assign an identifier to their own software, and situations where people who did not create the software (non-developers) may register an identifier for it in order to cite it. For non-developers, archiving a piece of software may not be possible. It is possible though for non-developers to register an identifier for software using a registry such as [SciCrunch](#) (RRIDs) or [ASCL](#) (ASCL IDs).The following are just a few examples of situations that may be cause for non-developers to create an identifier for a piece of software:
- The software is commercial or proprietary
- The software may be legacy code without clear authors and the full source cannot be located.
- The software may be a combination open and closed software without clear authors (e.g. a pipeline like [spec2d](#)).

We want to avoid creating new identifiers for software that already has a primary identifier (e.g., a DOI) where it is not possible to create and maintain a semantic relationship between the identifiers, because this makes it harder to reconcile identifiers for purposes of assigning credit and determining provenance, and it is still a challenge to know where to find metadata for software in order to appropriately describe it in a registry record.

Dan, Daina, and Neil wrote a detailed [blog post about this topic](#).

## 6.6. Bibtex

Based on discussion in Section 3.4.

- There is a bibtex software type, but a lot more is needed around it to make it truly useful.
- For example, the DataCite Content Resolver in Section 3.4 returns a bibtex item of type @misc for software, and this seems likely to be the case for other tools as well.
- Also, there are needs around how a bibtex entry of type @software is processed. Currently, there is no specific driver set up in standard.bbx, it uses the one defined for @misc. See https://github.com/force11/force11-sciwg/issues/48#issuecomment-371871401 for more info.
- Journal and conference-specific class files might also need to be updated to be able to properly process @software entries

## 6.7. Text citation styles for software

Based on discussion in Section 3.5 and Section 4.4.1

- How can software be identified as software in text citations when the style guides do not include it?
- If JATS included "software" and related terms (which might be recommended by JATS4R), it would be easier for text citation styles for software to be standardized.

## 6.8. Generating DOIs for "unversioned" software "packages"/"works"

Based on discussion in Section 3.5

- There is currently no public way to directly obtain a DOI for "unversioned" software "packages"/"works".
- This can be done as a byproduct of depositing software with Zenodo, as depositing the first version will create a DOI for that specific version and a DOI for the set of versions that are created with Zenodo, but this is not the same as being able to do this generically.
- Organizations who register DataCite DOIs can do this via Fabrica, but this is a small number of organizations.

## 6.9. Discussion of specific tools

Based on discussion in many sections. These tools include Software Heritage, CFF, CodeMeta, cffconvert.

- Do we want to talk about specific tools? Doing so makes the document more useful, but also makes it likely to go out of data quickly.

Our decision is to try to talk at a high level, but also include specific tools as examples, and to emphasize particular tools/standards that we think should be adopted.

## 6.10. Out of scope items

- Transitive credit. Citing software directly in a paper or product is analogous to traditional citation of scholarly works, but software is much more frequently 'cited' via dependency chaining and provenance tracking. Katz & Smith discuss transitive citation and the potential impact that would have on our understanding of the impact of software on scholarly and other endeavours. Both dependency chains and provenance graphs provide mechanisms for assigning credit for an impactful outcome such as a manuscript publication to the data and software on which the findings were based. Citations to a downstream package can then be properly attributed to upstream software and data that were fundamental to the cited outcomes, even though only the proximate package was directly cited in the reference list of the paper. While this concept is a valuable idea in discussion of credit, it is out of scope in a discussion of citation.
- Review of software is out of scope in a discussion of citation, as citation indicates that software has been used, whether or not it has been reviewed. Similarly, review of publications and review of proposals/grants are out of scope, except that the processes should have policies that ensure software citation is implemented and used within them.
- Layers of software re Hinsen. This is mostly subsumed in the Section 2 discussion of software types.

## 6.11. Registries that store software metadata

As indicated in Section 3.3, there are at least 2 registries that store metadata for items that include closed source software as well as software for which metadata were not created by the authors of the software: SciCrunch and ASCL. Records in these registries do not necessarily enable native software citation, but they allow a user to cite software indirectly by pointing to associated documentation and metadata (e.g. user manual, published paper). Citation of a registry record may be appropriate in cases where native software citation is not possible and metadata is publically available (e.g. CSC, CSUV). Registry records should not be cited in favor of published software records (e.g. Zenodo).

Open issues for these repositories include how they work together to:
1. provide common services
2. define standards of how these are maintained
3. not duplicate each other's identifiers

Further work is needed here, which is being considered by the FORCE11 Software Citation Implementation Group's Repository Best Practices Taskforce.

## 6.12. Collections: frameworks, components, and pipelines

Some software is developed as components for specific frameworks. This is briefly mentioned in Section 2. However, how this software should be cited is not clear. First, a user of a framework may not know which components are actually being used, as they may only see the framework itself. If the components can be identified, then all the issues in Section 3 regarding metadata arise, e.g. how is the component identified, who are its authors, etc. Pipelines or frameworks may also be cited both as concepts as well as specific version. In addition, Research Object can be collections of components, some of which may be software. Examples of this issue occur may occur with Python and R, and while R has provided a solution, it does not completely match the guidance in this document (see unresolved issues in Section 3.3).

## 6.13. Identifiers for concepts vs for collections of versions

Based on discussion in Sections 2 and 3.1

Identifiers for concepts is an idea that has multiple possible meanings. One idea is that the identifier is for the overall concept: the abstract idea of the software that is also related to all versions of the software (instantiations of the concept). A second idea is that the identifier is for a collection of versions of the software, which could be some or all of the versions. For example, when a version of software is first published to Zenodo, Zenodo creates two identifiers, one for this and any future versions (i.e., a collection ID), and one for this specific version. If a later version of this software is then published to figshare, the Zenodo ID for the collection cannot easily be updated to include that non-Zenodo version; thus the Zenodo ID is not an identifier for the software concept. (Note that in theory, it could be, but in practice, we don't have the publicly-available tooling to make this work.)